# All–optical sub–diffraction multilevel data encoding onto azo-polymeric thin films.

*Matteo Savoini, Paolo Biagioni, Lamberto Duò, and Marco Finazzi.\**

LNESS-Dipartimento di Fisica, Politecnico di Milano, Piazza Leonardo da Vinci 32, 20133 Milano, Italy.

\* marco.finazzi@fisi.polimi.it; tel +39 02 2399 6177; fax +39 02 2399 6126.

By exploiting photo-induced reorientation in azo-polymer thin films, we demonstrate all-optical polarization-encoded information storage with a scanning near-field optical microscope. In the writing routine, 5-level bits are created by associating different bit values to different birefringence directions, induced in the polymer after illumination with linearly polarized light. The reading routine is then performed by implementing polarization-modulation techniques on the same near-field microscope, in order to measure the encoded birefringence direction.

Since the beginning of the computer era, data storage has been a fundamental issue in the scientific world. During the last decades, considerable efforts have been devoted to find solutions other than magnetic storage, allowing for cheaper substrates, smaller bit definition and lower power consumption during encoding and decoding.[1] In this frame, polymeric materials have been extensively exploited, especially those, such as azo-polymers, that present photo-induced motion and reorientation.[2] When illuminated with an appropriate linearly-polarized source, in fact, these polymers undergo isomerization cycles which eventually orient them with respect to the polarization direction.[3] This characteristic has attracted the attention of the scientific community, which has devoted many efforts to investigate the applications of such polymers to far-field data encoding,[4,5,6] optical holography,[7] multipolar nonlinear optical memories,[8] 3D information storage,[9] and near-field binary encoding.[10] With an increasing exposure energy, azo-polymers also present a polarization dependent macroscopic translational movement. This characteristic has been exploited for applications in nano-movement,[11] nano-fabrication,[12] visualization of local field enhancements due to plasmonic resonances,[13] and mapping of field components.[14]

Initially, azo-polymers were used in the liquid crystal form,[5] but Nathanson *et al.* have demonstrated that liquid crystallinity is neither a necessary nor a best suited condition for optical storage.[4] Liquid crystal azo-polymers, in fact, present glass transition temperatures comparable with room temperature, leading to a lower storage stability since the ordering tends to smear out in ambient conditions.[3]

In this paper, we demonstrate the possibility to exploit optically-induced orientation of azo-polymers for all-optical, high-density multilevel data encoding and we define a new operative method for the development of organic-based information storage technology. We optimize photon energies and fluencies for the writing and reading routines and we implement near-field scanning optical microscopy (NSOM)[15] with polarization modulation and analysis in order to overcome the diffraction limit for optical resolution. In this respect, the polarization preserving feature of the setup is a key point to reliably control information encoding and decoding. We thus achieve efficient writing/reading/rewriting of 5-level bits (pentabits), with a lateral size of about 250 nm.

We use a commercially available azo-moiety Disperse Red-1 blended in a poly(methylmethacrylate) (PMMA) matrix with a molar concentration of 25%. This blend presents a rather high glass transition temperature of 102 °C. A sample thickness of about 200 nm is chosen since, for this film thickness, the optical response (i.e. the microscopic reorientation) of the polymers reaches its asymptotical value, as reported in the literature.[16]

For near-field encoding/decoding, we use a commercially available NSOM (AlphaSNOM, WITec GmbH) modified in order to accommodate the optical elements presented in Figure 1.

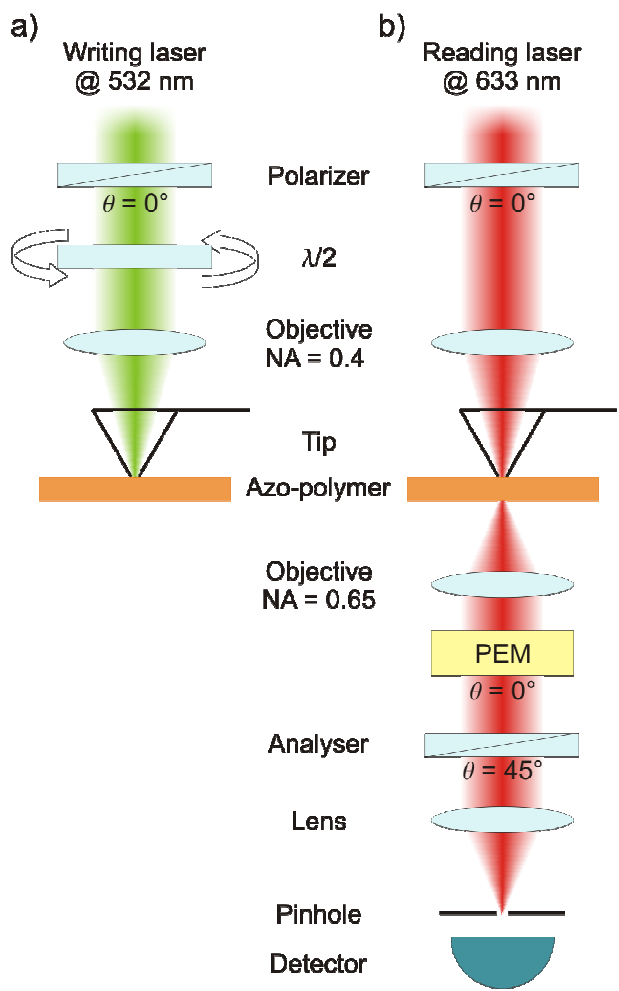

**Figure 1.** a) Encoding set up: the birefringence axis induced in the azo-polymer film is determined by the orientation of the half-wave plate. b) Decoding set up to measure the induced birefringence. The polarizers, the photo-elastic modulator (PEM), and the analyzer have their optical axes at fixed angles $\theta$ with respect to the linearly polarized electric field of the light.

The instrument employs a hollow-pyramid probe with an optical resolution comparable with the tip aperture nominal diameter, equal to 120 nm. As recently demonstrated, this type of near-field probes ensure preservation of the incoming light linear polarization.[17]

The encoding routine is performed at a light wavelength $\lambda = 532$ nm (corresponding to the emission of a frequency-doubled Nd:YAG laser), at which the film absorbance is relatively high (about 12%). On the other hand, a low absorbance is needed during the decoding routine, otherwise the stored information might be corrupted. In this case we employ a wavelength equal to 633 nm (supplied by a He-Ne laser), for which the measured absorbance is less than 1%. The information is encoded by optically inducing birefringence over small sub-wavelength volumes. This mechanism involves photochemical excitation of the azobenzene group, activated with linearly polarized light. In this excitation process, the azobenzene unit undergoes a trans-cis-trans photo-isomerization cycle, which usually takes about $10^{-9}$ s. Photo-isomerisation is a reversible process in which the single azo-unit is switched between a ground state (trans) and a more energetic one (cis). Due to thermal relaxation, the molecules then relax back to the ground state, in random orientation. However, those relaxed molecules that fall perpendicular to the incoming light polarization are no longer able to absorb energy, and remain fixed. Thus, on a time scale of the order of $10^{-4}$ s, there is a statistical enrichment of azo-mers perpendicular to the linearly-polarized light electric field.[4] This induces a birefringence axis in the sample that can be optically measured.

Operationally, the pentabit-encoding routine is performed by illuminating the sample surface with an appropriate polarization, keeping the tip in close proximity to the sample at the writing position. In the decoding routine, instead, the tip is raster scanned on the encoded pentabit. The light transmitted through the film is collected by a microscope objective, crosses a photo-elastic modulator and is measured with a lock-in amplifier connected to the detector, yielding an amplitude

$$A \propto \delta \sin(2\phi), \qquad (1)$$

where $\phi$ is the angle between the direction of the induced birefringence axis and the electric field of the decoding polarized light. The parameter $\delta$ stands for the birefringence induced in the sample and is defined as

$$\delta = 2\pi \frac{d}{\lambda} \Delta n, \qquad (2)$$

with $\lambda$ being the wavelength of the decoding light, $d$ the sample thickness, and $\Delta n$ the difference in the refractive index experienced by the two light components parallel and perpendicular to the birefringence extraordinary axis.

We have experimentally verified that, on the average, $A \neq 0$ even in non-encoded areas. This is due to the birefringence introduced by the optical system and/or by the microscope glass supporting the film. Although this might result in a loss of sensitivity, we can define levels that are both positive and negative compared to this offset. This is possible since the $\phi$-dependence shown by the amplitude signal is sign sensitive.

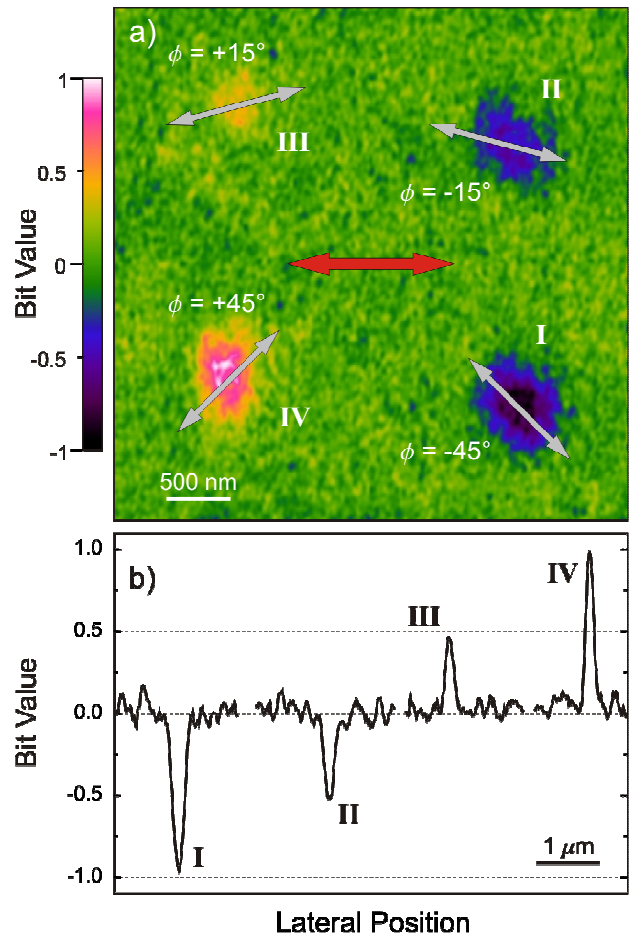

**Figure 2.** a) Birefringence map of a sample region containing four different pentabits. A constant background has been subtracted. The grey arrows represent the different orientations of the birefringence axis associated to each pentabit, while the red arrow indicates the decoding polarization direction. The values of the angle $\phi$ between the green and red arrows are also indicated. b) Line profile of the pentabits: four well-defined levels (1, 0.5, -0.5, -1) are visible. The fifth level (0) is represented by the signal associated to non encoded areas. The FWHM of each pentabit is 250 ± 15 nm.

Figure 2a shows a birefringence map where four different bits have been encoded with a power impinging into the tip equal to 1 μW. From now on, the reported powers are intended to be the ones incident into the tip. The measured far-field throughput of our NSOM tips is typically $5 \times 10^{-3}$ for $\lambda = 532$ nm. The encoding procedure has been performed with light presenting a linear polarization along different orientations: $\phi = \pm 45°$ and $\phi = \pm 15°$ with respect to the decoding polarization direction. We choose such angles to have equally-spaced 5-level encoding, in agreement with Eq.1, the 0 (zero) level corresponding to no encoding or, equivalently, to $\phi = 0$. The differences between the pentabits are not due to a different exposure energy (defined as input power × exposure time) but strictly to the $\phi$-dependence of this encoding routine. Figure 2 therefore demonstrates the possibility offered by our polarization maintaining setup to reliably write and retrieve different units of information in a 5-level system of encoding by controlling the polarization of the near-field component of the light. The encoded pentabits show a long lasting persistence,[3] which we verified by measuring the sample after one month storage at ambient conditions, revealing no appreciable degradation of the stored information.

In Figure 2 the exposure time is 30 s. We chose such a long exposure time since it represents a good compromise between encoding speed and lateral definition of the pentabits. High fluencies, in fact, allow writing the information in much shorter times, but also tend to increase the temperature in the exposed area. This leads to the thermally activated motion of the azo-moieties, resulting in larger bit areas and a worse signal-to-noise ratio,[18] as discussed below.

The decoding routine has been performed on an area of $4 \times 4$ μm$^2$, with a point integration time of 10 ms and an incident laser power of about 50 μW. These values have shown to be a good compromise between reading speed and signal-to-noise ratio. The topography and phase signals (not shown) collected during the decoding step are flat within the instrument noise, demonstrating that, with such a low encoding exposure energy, no macroscopic movement is activated. The phase of the decoded signal turns out to be dominated by the residual birefringence of the optical elements since no contribution is observed in connection with the different pentabits.

All four levels present a lateral dimension with FWHM = 250 ± 15 nm. We would like to stress that the size of each pentabit results from the convolution between the lateral size of the encoded area and the resolution associated with the decoding step. Note that this size is below the FWHM lowest limit one could expect by combining diffraction-limited encoding and decoding routines, which can be estimated around 310 nm (assuming a numerical aperture equal to 1.4).

With a lateral optical resolution associated to the decoding step equal to 120 nm, we can estimate that the polymer area reoriented during the encoding step has a FWHM diameter of about 220 nm. This area is larger than the one directly exposed to the tip aperture during the encoding routine. The broadening effect is likely due to the fact that each azo-moiety is coupled to the neighbouring ones because of short-range interactions.[19,20]

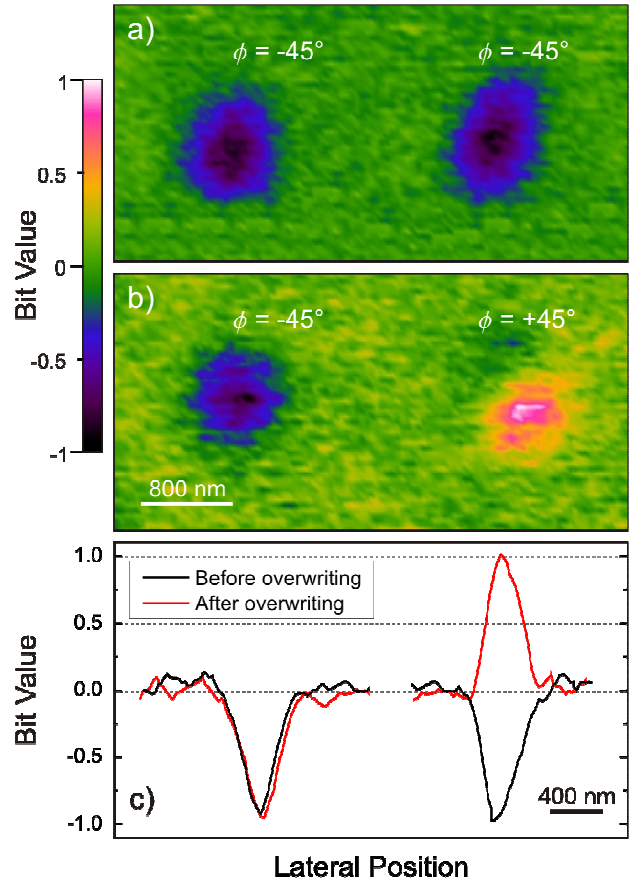

**Figure 3.** Overwriting the information: the pentabit on the right in a) is re-encoded with the opposite value in b), while the left one is maintained as a reference; $\phi$ is the angle between the encoded birefringence axis and the decoding polarization direction. c) Pentabit line profile of the pentabits before and after the overwriting procedure.

Our system for data encoding also provides a reliable overwriting procedure. In our case, changing the information associated to a pentabit consists in directly writing a new pentabit in the same position, with the advantage that the re-orientation of the azo-moieties does not present any hysteretic behaviour, at variance with magnetic storage supports. Figure 3 shows the overwriting of a -1 bit into a +1 bit ($\phi$ = -45° is changed into $\phi$ = +45°). This is obtained with the same writing procedure and the same conditions employed to obtain Figure 2.

As previously stated, the lateral dimension of the pentabits is affected by parameters such as the exposure time and/or energy. Figure 4 shows two pentabits encoded with different exposure times but the same exposure energy as the one employed to obtain Figure 2. The first pentabit is encoded with an exposure time $T_{exp} = 1$ s and an incident power equal to 30 μW, while the second is encoded under the same conditions as in Figure 2 ($T_{exp} = 30$ s, incident power equal to 1 μW). As it can be noted, a higher power (corresponding to a shorter exposure time) leads to a higher FWHM lateral size (300 nm for $T_{exp} = 1$ s vs. 250 nm for $T_{exp} = 30$ s).

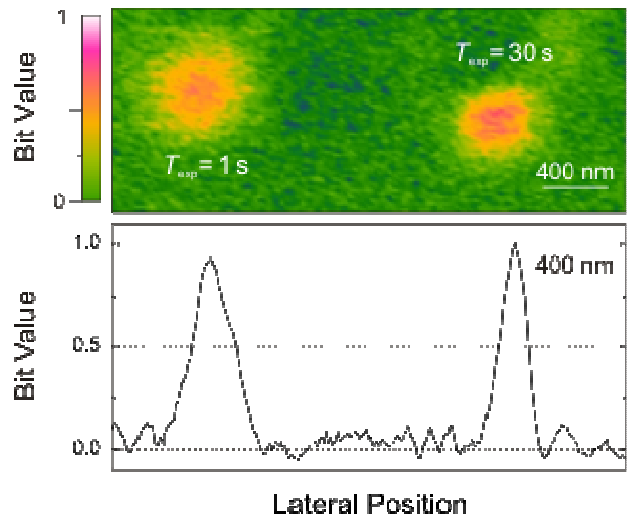

**Figure 4.** a) Amplitude signals of two nominally equally pentabits ($\phi = +45°$ for both pentabits) encoded with the same total exposure energy (30 μW·s), but different exposition times $T_{exp}$. b) Line profile of the two pentabits. The FWHM of the pentabits is 300 nm (left) and 250 nm (right), respectively.

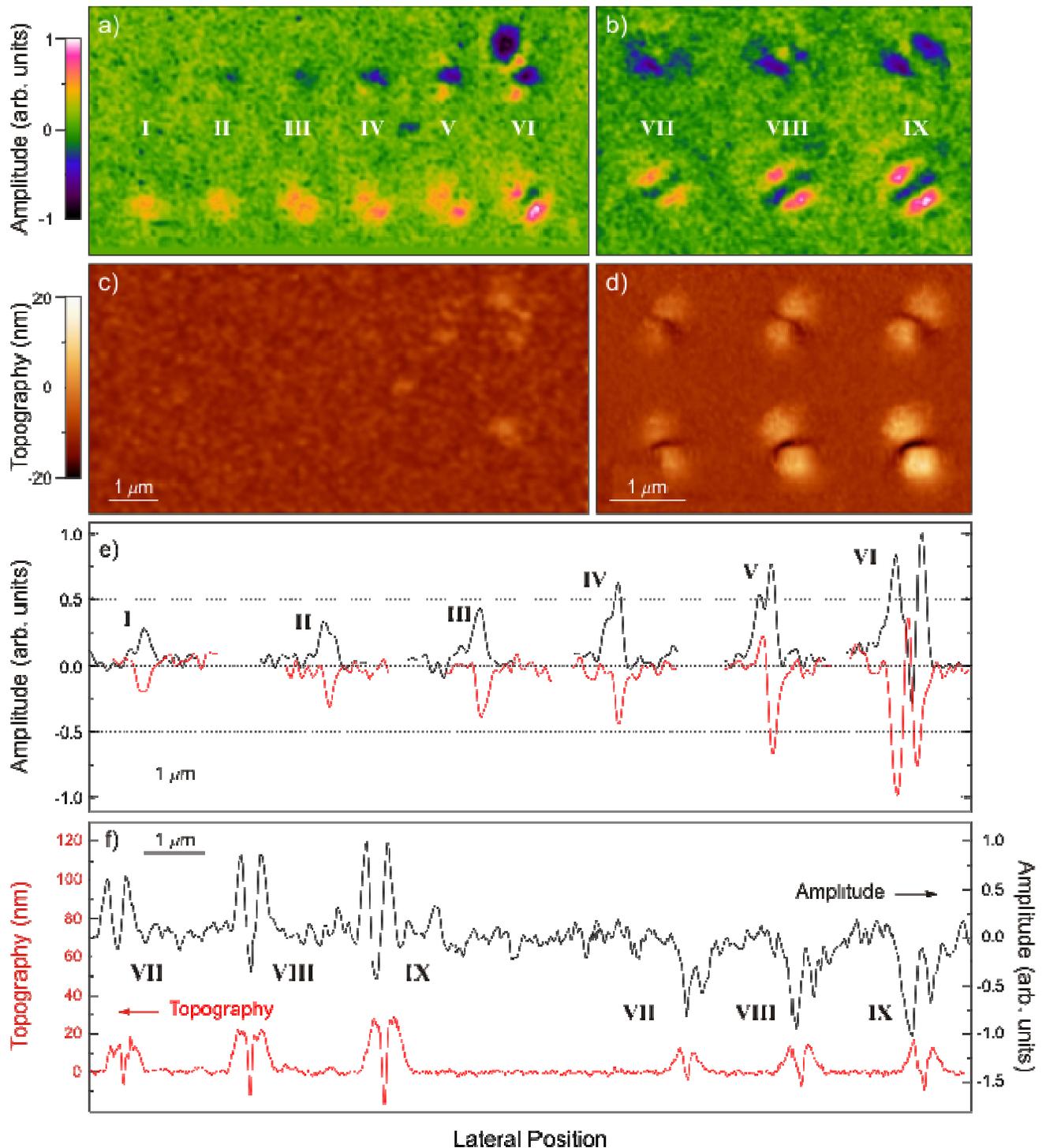

**Figure 5.** Effects of the total exposure energy on the azo-polymer film. a,c) Birefringence map and corresponding topographic signal for couples of pentabits ($\phi = \pm 45°$) as a function of the exposure time for a fixed incident power equal to 15 µW. From I to IV the exposure time is doubled at each step, starting with 1 s for I. The corrensponding amplitude line profile is presented in e). b,d) Birefringence map and corresponding topographic signal for couples of pentabits ($\phi = \pm 45°$) as a function of the exposure time for a fixed incident power equal to 1.5 mW. The exposure time is 1 s, 2 s, and 4 s for VII, VIII, and IX, respectively. f) Topography and amplitude profile of the pentabits in b,d).

A final issue, shown in Figure 5, is the importance of the total exposure energy: too low an energy induces only a partial reorientation of the azo-moieties, leading to blurry signals, as shown in Figure 5a). Increasing the exposure energy (from left to right in Figure 5a,b), a more uniform and well defined orientation is induced in the exposed area and the associated signal-to-noise ratio improves in the decoded amplitude. On the other hand, too high an exposure can lead to the macroscopic movement of the azo-chain. The result is the presence of undesired artefacts in the decoding maps, due to the interaction of the tip near field with the inhomogeneities caused by the macroscopic movement itself (see pentabits VI in Figure 5a,c). These effects are even more evident in the couples from VII to IX, which show "butterfly shapes" both in the amplitude and in the topographic maps (Figure 5b and Figure 5d, respectively) when excessive exposure energies are used. In this case, even the topography remarkably retains a polarization dependence. The demodulated amplitude signal $A$ presents two wing-like lobes divided by a central line, due to the dip formed by the macroscopic motion of the azo-moieties involved in the photochemical excitation process. Upon increasing the exposure energy, more material is accumulated into the side lobes, while a depletion up to 40 nm is formed in the middle (see Figure 5f).

In conclusion, we have demonstrated long lasting all-optical multilevel information storage with sub-diffraction resolution, based on polarization encoding and decoding in azo-polymer thin films. Pentabits with a diameter of roughly 250 nm have been written, read and re-written by a proper choice of photon energies, fluencies and polarization. In perspective, encoding on more than 5 levels is possible, the only limitation being the sensitivity of the decoding set up. Nano-patterning the surface of the films to form a grid with squares having dimensions comparable with the tip diameter could possibly be an efficient solution to avoid the smearing of the encoded information over an area larger than the tip aperture. In such a way the resolution and density of the bits could possibly be highly increased.

The authors acknowledge G. Brusaferri for assistance during the experimental sessions. All images have been elaborated using WSxM.[21]